\documentclass[aps,superscriptaddress,preprint]{revtex4-2}

\usepackage{graphicx}
\usepackage{epstopdf}
\usepackage{xcolor}
\usepackage{color}
\usepackage{dcolumn}
\usepackage{bm}
\usepackage{hyperref}
\definecolor{dblue}{RGB}{46,48,146}
\hypersetup{
hidelinks,
colorlinks=false,
linkcolor=dblue,
citecolor=dblue,
urlcolor=black
}

\usepackage{setspace}
\begin{document}

\title{The Magnetic Dislocation in Photonics}

\author{Danying Yu}
\affiliation{State Key Laboratory of Advanced Optical Communication Systems and Networks, School of Physics and Astronomy, Shanghai Jiao Tong University, Shanghai, 200240, China}

\author{Kun Ding}
\email{kunding@fudan.edu.cn}
\affiliation{Department of Physics, State Key Laboratory of Surface Physics, and Key Laboratory of Micro and Nano Photonic Structures (Ministry of Education), Fudan University, Shanghai 200438, China}

\author{Xianfeng Chen}
\affiliation{State Key Laboratory of Advanced Optical Communication Systems and Networks, School of Physics and Astronomy, Shanghai Jiao Tong University, Shanghai, 200240, China}
\affiliation{Shanghai Research Center for Quantum Sciences, Shanghai, 201315, China}
\affiliation{Collaborative Innovation Center of Light Manipulations and Applications, Shandong Normal University, Jinan, 250358, China}

\author{Luqi Yuan}
\email{yuanluqi@sjtu.edu.cn}
\affiliation{State Key Laboratory of Advanced Optical Communication Systems and Networks, School of Physics and Astronomy, Shanghai Jiao Tong University, Shanghai, 200240, China}

\begin{abstract}
The dislocation created in the topological material lays the foundation of many significant findings to control light but requires delicate fabrication of the material. To extend its flexibility and reconfigurability, we propose the magnetic dislocation concept and unveil its properties in a representative model, which effectively combines the topological defect and edge mode at the magnetic domain wall. The results include distinct localization modes and robust light trapping phenomena with the rainbow feature where the eigen-energy of each light-trapping state can be linearly tuned by the magnetic dislocation. The conversion from the trapping state to edge modes can be harnessed by further adiabatically pumping light across an amount of the magnitude of the magnetic dislocation. Our work solves a fundamental problem by introducing magnetic dislocation with new light-manipulation flexibility, which may be implemented in a variety of platforms in photonic, acoustics, and optomechanics with dynamic modulations and synthetic dimensions.
\end{abstract}

\maketitle

\section{Introduction}

Dislocation, as the topological defect, describes the distortion of the hosting lattice, usually characterized as the real-space topology that cannot be eliminated by local continuous modifications \cite{1,2,3,4}. Recently, it has been realized that the interplay between dislocations and band topology in topological materials brings novel ways in light manipulations against disorders \cite{5,6,7,8,9,10,11,12,13,14,15,16,17,18,19,20,21}, including the light-trapping \cite{13}, bulk probes of higher-order topological insulators \cite{18}, simulating three-dimensional topology \cite{20,R20_1,R20_2,R20_3}, controlling the topology of material \cite{21}, and probing the topology of the system \cite{22,R22_5}. Previous studies have focused on the dislocation configurations from spatially designed topological defect \cite{4,5,13,21,R22_1,R22_2,R22_3,R22_4}, which has limited reconfigurability, particularly in the optical regime \cite{23}.

The effective magnetic field for photonics provides a powerful approach to control the motion of light in a flexible way \cite{24,25,26,27,28,29,30}, with which the photonic analogy of quantum Hall effect can be realized in photonics platforms \cite{25,26,27,28}. As for dislocation in the hosting lattice, it has been indicated that the effective gauge potential generated from the geometrical dislocation is the main reason affecting the property of the topological defect mode \cite{4}. It is, therefore, of fundamental interest to explore the direct dislocation in an effective magnetic field, namely \emph{magnetic dislocation}, in photonics, which not only deepens the fundamental understanding of the dislocation mechanism but also shows new potential in reconfigurable light manipulation.

In this work, we theoretically study the fundamental problem of the magnetic dislocation in a spatially uniform tight-binding (TB) lattice. The magnetic dislocation is introduced by designing non-continuously distributed hopping phases to construct a magnetic singularity in the underlying lattice without spatial defect, as shown in Fig.~\ref{Figure.1}(a). The introduction of such artificial magnetic dislocation is in-principle different from previous models with dislocation well-defined in spatial structures \cite{4,r57}, but is universal to be possibly constructed for waves using recent experimental technologies in photonics \cite{31,32,33}, acoustics \cite{34,35}, and optomechanics \cite{36,37,38}. We unveil such concept of magnetic dislocation as the composition of two effective mechanisms, namely the topological defect and edge mode at the magnetic domain wall simultaneously. Therefore, by tuning the discontinuous hopping phase, light can be trapped at the singularity of the magnetic flux with the rainbow feature \cite{39,40}, i.e., the eigen-energy of each light-trapping state varies with the tuning of the magnetic dislocation monotonically. Such light trapping state can be converted to edge modes by adiatically pumping the light. Our proposal can be generalized to all electromagnetic waves \cite{41}, which may trigger further research interest in topological defects with magnetic singularities and novel wave manipulations \cite{42,43,44}.

\section{Model}
\begin{figure}[htbp]
\centering
\includegraphics[width=0.9\textwidth ]{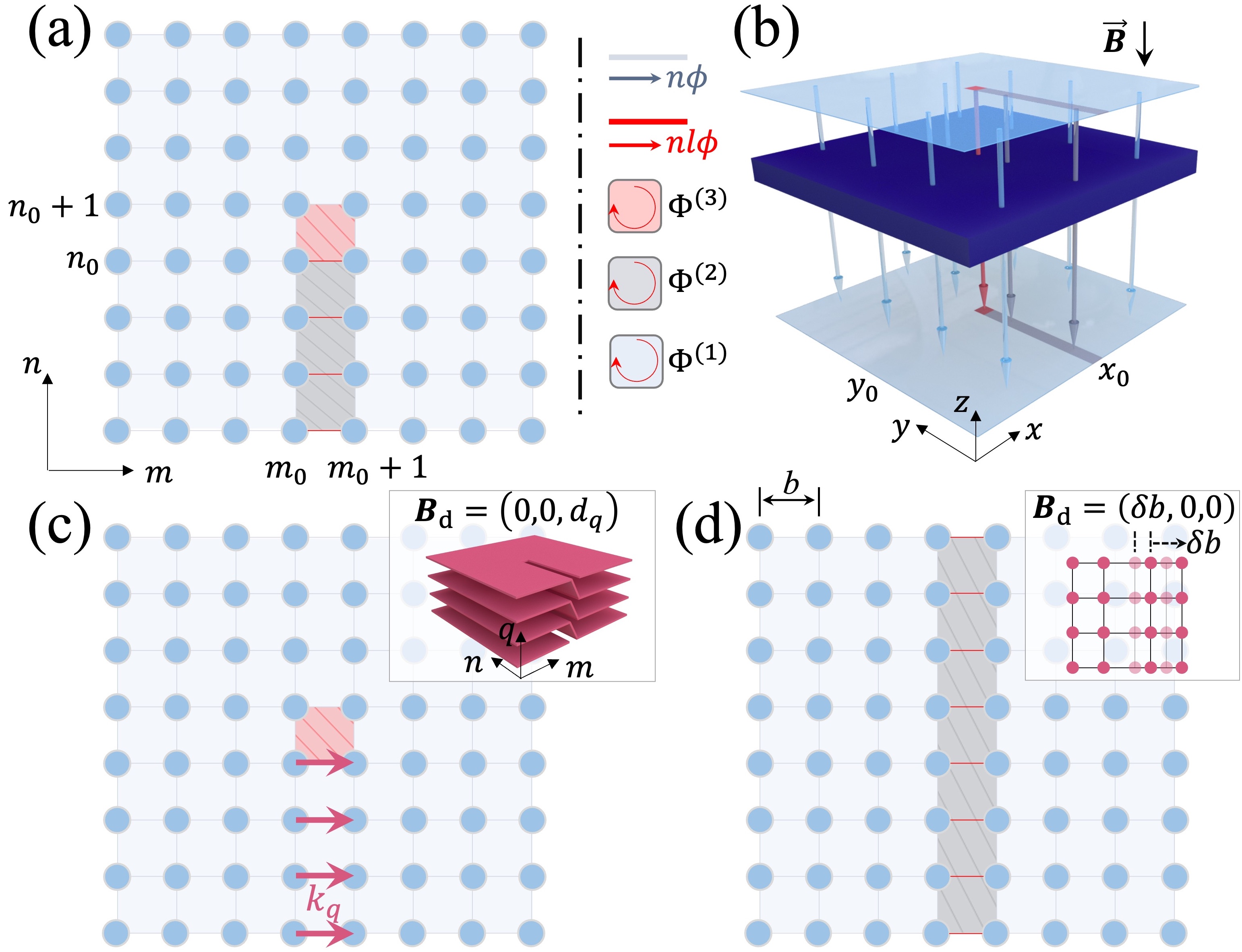}
\caption{The model of the magnetic dislocation. (a) The 2D lattice with the designed distribution of hopping phases. (b) The continuous model subjected to a non-uniform magnetic field \cite{54}. (c) The deformation along the $q$ dimension forms a 3D screw dislocation, which can be effectively characterized by deforming the magnetic field at the center (the magnetic singularity). (d) The deformation along the $m$ dimension forms the glide of the lattice sites, which can be effectively characterized by deforming the magnetic field in the middle (domain wall). Zoom-in figures label the corresponding Burger vectors to characterize deformations in (c) and (d).}\label{Figure.1}
\end{figure}

We consider a two-dimension (2D) topological lattice model including a uniform effective magnetic flux [see Fig.~\ref{Figure.1}]. The Landau gauge is taken with horizontal hopping phases on the $n$-th row as $n\phi$ \cite{45}, which supports a uniform effective magnetic field $\textbf{\emph{B}}_\mathrm{eff}=\frac{1}{b^2}\int_\mathrm{plaquette}\textbf{\emph{A}}_\mathrm{eff}d\textbf{\emph{r}}=\frac{\phi}{b^2}$ \cite{25}, with $\textbf{\emph{A}}_\mathrm{eff}$ being the effective gauge potential and $b$ being the lattice constant. We take $\phi=\pi/2$ throughout this paper. The corresponding Hamiltonian is
\begin{equation}\label{1}
 H=\sum_{m,n}\kappa a_{m,n}^\dagger a_{m+1,n} e^{-in\phi}+\kappa a_{m,n}^\dagger a_{m,n+1}+h.c.,
\end{equation}
where $a_{m,n}^\dagger (a_{m,n})$ is the creation (annihilation) operator and $\kappa$ is the coupling strength. Like the quantum Hall effect in the electronic system, the effective magnetic field imposed on photons makes the system topologically non-trivial with edge states \cite{27}.

On the basis of this conventional model (\ref{1}), we introduce the dislocation directly in the magnetic field. As illustrated in Fig.~\ref{Figure.1}(a), we modify hopping phases between $m_0$-th and $(m_0+1)$-th columns from $n\phi$ to $nl\phi$ for those horizontal hoppings at $n\leq n_0$, where the real number $l$ is a parameter that brings the discontinuity and determines the magnitude of the magnetic dislocation (while the magnetic field is uniform if $l=1$). $n_0$ labels a fixed position. Such magnetic dislocation is designed refer to the counterpart of edge dislocation in spatial geometry \cite{13,14,18,22,r58,r59}. Such design gives three types of magnetic fluxes inside the lattice, where the one in the middle plaquette in Fig.~\ref{Figure.1}(a) is $\Phi^{(3)}=(n_0+1)\phi-n_0 l\phi$, fluxes in each plaquette underneath is $\Phi^{(2)}=l\phi$, and all rest gives $\Phi^{(1)}=\phi$. Such introduced magnetic dislocation may be linked to the continuous model with the magnetic field supporting the magnetic singularity in the middle and corresponding line discontinuity (see Fig.~\ref{Figure.1}(b) with details in supplementary material \cite{54}).

Such discontinuous hopping phase distribution can promisingly be fulfilled in photonics with tailored designs \cite{46}. For example, one can consider a synthetic frequency lattice  built in coupled ring systems under dynamic modulations that supports hopping phase $n\phi$ along the frequency axis of light \cite{27,33,47,48,49,50,51,52,53}. By adding auxiliary ring to split resonant frequency mode and then modulating the frequency between split supermodes with designed modulation phase, one can construct the effective hopping phase $nl\phi$. Therefore, in this photonics platform, the parameter $l$ can be tuned by varying the amplitude of the modulation phase between split supermodes (see supplementary material for more details \cite{54}). Besides, other implementations may also be proposed in other photonic platforms supporting effective gauge fields \cite{25,26,32,55}, or with acoustics \cite{56} and optomechanics \cite{37}. Moreover, although our model is based on the distribution of hopping phases, the concept of magnetic dislocation may be extended to other models.

We find the model of magnetic dislocation can be considered as the composition of two effective mechanisms with different Burger vectors, i.e., the singularity of the flux $\Phi^{(3)}$ is equivalent to the topological-defect model in Fig.~\ref{Figure.1}(c) while the line discontinuity of the flux $\Phi^{(2)}$ corresponds to a magnetic domain wall described in Fig.~\ref{Figure.1}(d). The former model supports the topological defect as $-\Phi^{(3)}+\phi$, mapping to an effective momentum $k_q$ which links to a three-dimensional lattice with screw dislocation  and gives the Burger vector $B_\mathrm{d}=(0,0,d_q)$ \cite{54}, where $d_q$ is the lattice constant along the $q$ dimension. The latter one supports edge modes from breaking time-reversal symmetry at the middle magnetic domain wall, which supports $B_\mathrm{d}=(\delta b,0,0)$, with $\delta b = (l-1)b$ from the magnetic deformation \cite{54}. Note that the magnetic dislocation in Fig.~\ref{Figure.1}(a) only supports half of the effective domain-wall model in Fig.~\ref{Figure.1}(d).

\section{Results}
The Hamiltonian describing the lattice with the magnetic dislocation in Fig. 1(a) reads,
\begin{equation}\label{4}
H_d=\sum_{m=m_0, \atop n\leq n_0}(\kappa a_{m,n}^{\dagger} a_{m+1,n}e^{-inl\phi}+\kappa a_{m,n}^{\dagger}a_{m,n+1}+h.c.) +\sum_\mathrm{others}(\kappa a_{m,n}^\dagger a_{m+1,n}e^{-in\phi}+\kappa a_{m,n}^\dagger a_{m,n+1}+h.c.).
\end{equation}
We set the lattice with $m\in[1,40]$, $n\in[1,40]$, $m_0=20$, $n_0=20$, and apply the periodic boundary condition (PBC) in both dimensions to study spectra. We first select $l=0.15$ to showcase different features from the magnetic dislocation.

\begin{figure}[htbp]
\centering
\includegraphics[width=0.9\textwidth ]{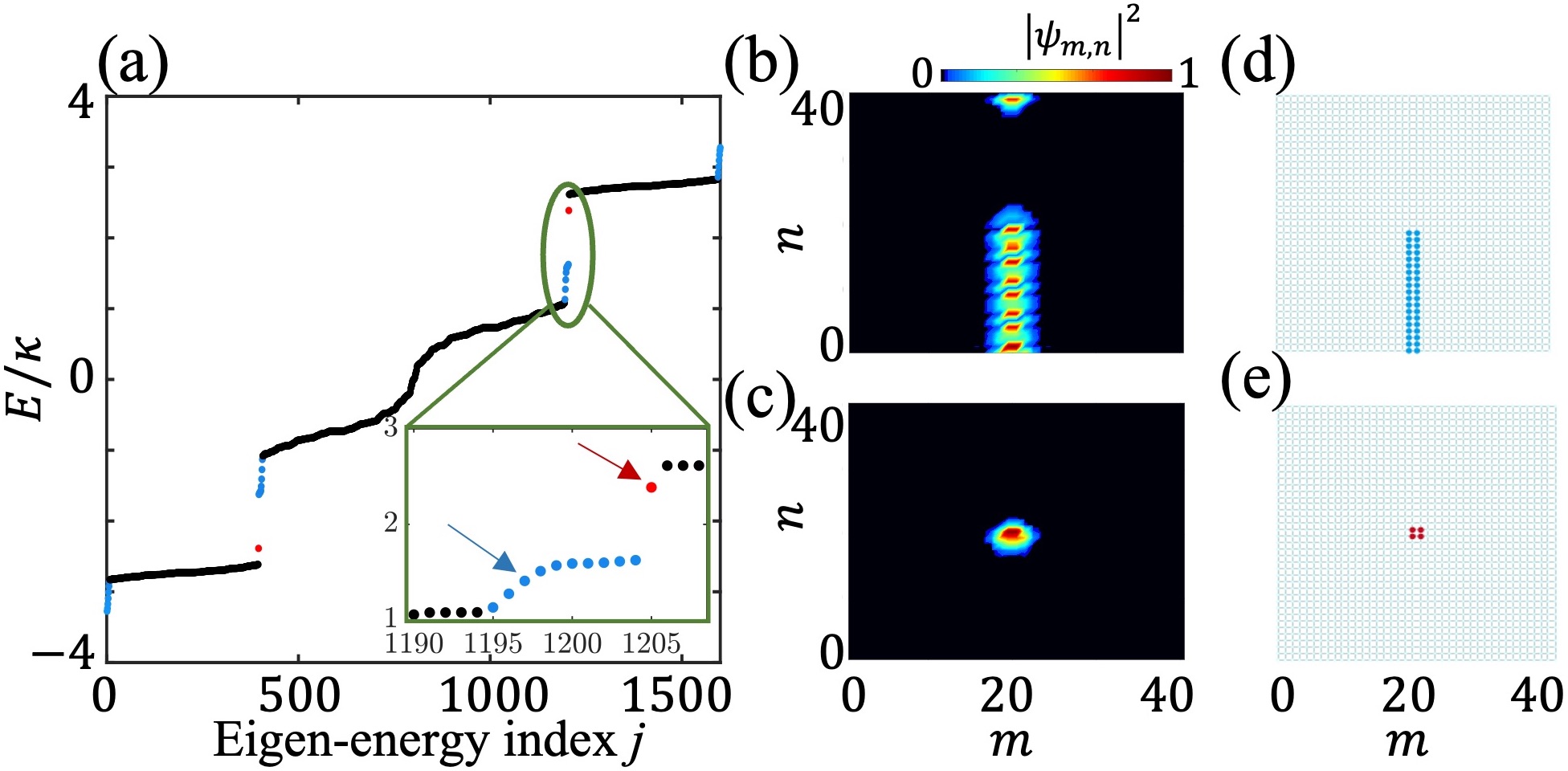}
\caption{Spectra and intensity distributions of eigen-states. (a) Spectra for $l=0.15$. Zoom-in plots are presented in green boxes. (b)-(c) Intensity distributions of eigen-states for the eigen-energies labeled by blue and red arrows in (a). (d)-(e) Schematic diagrams of standard line and point localization
modes, respectively. }\label{Figure.2}
\end{figure}

In the plotted spectrum in Fig.~\ref{Figure.2}(a), the immediate observation lies on the existence of multiple eigen-modes inside bulk gaps open by the background flux $\Phi^{(1)}$. We choose one mode at  $E=1.4\kappa$ and plot the eigen-state distribution in Fig.~\ref{Figure.2}(b), which shows the localization feature at lattice sites $(m,n)$ with $m=20$ or $21$ and $n\leq n_0$. Such eigen-state distribution is from the line discontinuity from the magnetic dislocation, dubbed as \emph{line localization mode}.

In Fig.~\ref{Figure.2}(c), we take $E=2.39\kappa$, and plot the eigen-state distribution. One notes that  it exhibits strong localization near the magnetic singularity $(20,20)$ that gives \emph{point localization mode}. Interestingly, these localization effects hold stable when the lattice size is increasing \cite{54}.

As a side note, the PBC we take here can induce the artificial discontinuity of the hopping phases along columns at $n=1$ and $n=40$. The eigen-state distribution in Fig.~\ref{Figure.2}(b) extends to the vicinity of top two middle sites. Such extension disappears once we consider a finite lattice model with open boundary condition (OBC)  in following simulations.

To quantitatively distinguish bulk modes and localization modes in the spectrum, we define the criterion using the inverse participation ratio (IPR) to characterize the mode feature, i.e., $\mathrm{IPR}\equiv\left[\sum_{m,n}|\psi_{m,n}|^4\right]/\left[\sum_{m,n}|\psi_{m,n}|^2\right]^2$ with $\psi_{m,n}$ being the eigen-state amplitude.

A standard point localization mode shown in Fig.~\ref{Figure.2}(e) has $\mathrm{IPR}=0.25$, refer the localization in the vicinity of the position of the singularity. For a standard line localization mode in Fig.~\ref{Figure.2}(d), $\mathrm{IPR}=0.025$ as the distribution in the lower half part extends among $n\in[1,20]$ in the middle of the $m$-axis. As for a typical bulk mode with distribution equally spread out the whole lattice, $\mathrm{IPR}=6.25\times 10^{-4}$.

\begin{figure}[htbp]
\centering
\includegraphics[width=0.9\textwidth ]{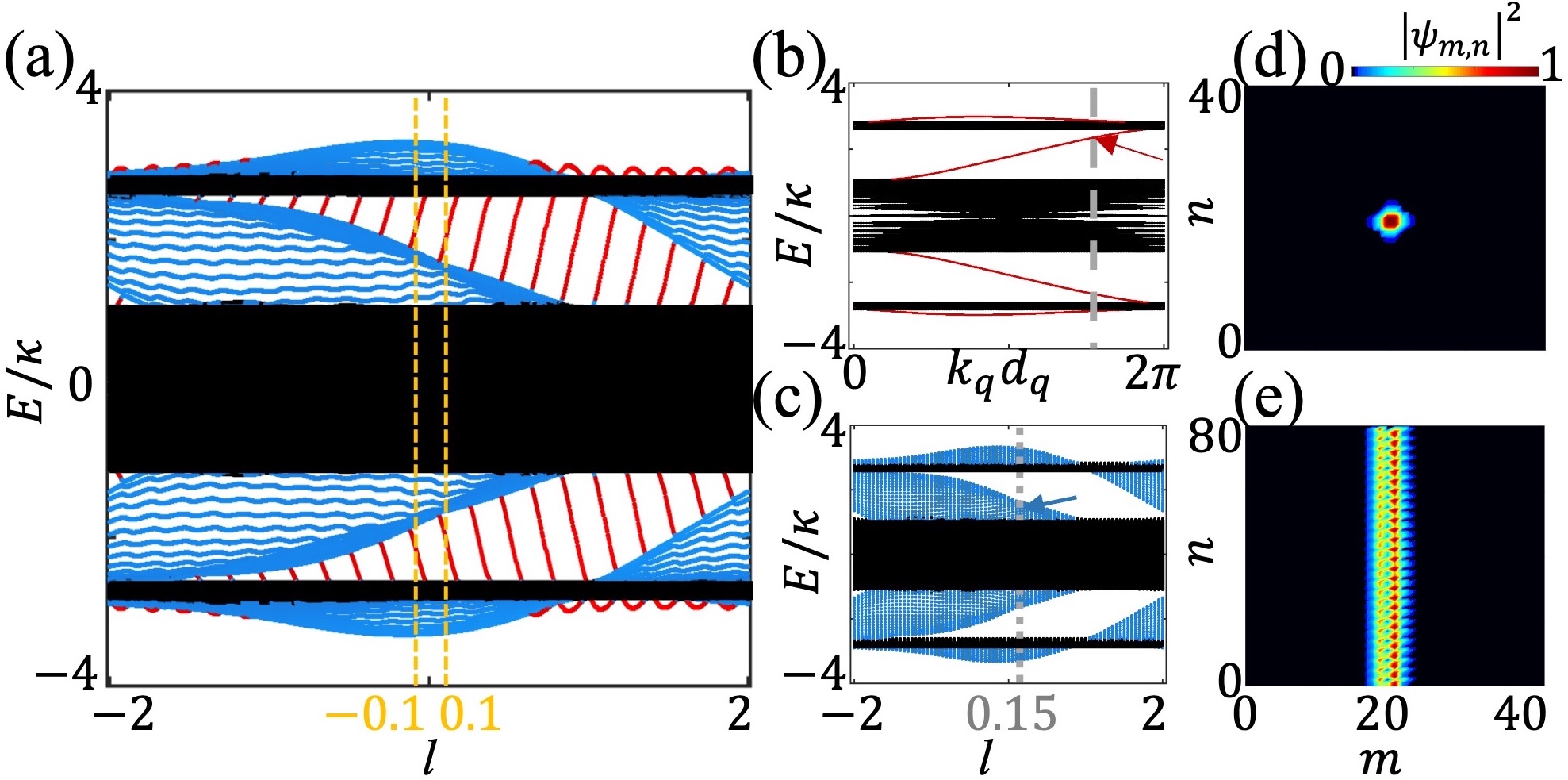}
\caption{The projected spectrum. (a) The projected spectrum along parameter $l$, where bulk states, line localization, and point localization are characterized as $\mathrm{IPR}<0.01$ (black), $0.01\leq\mathrm{IPR}<0.04$ (blue), and $\mathrm{IPR}\geq 0.04$ (red), respectively. (b)-(c) The spectrum from effective models in Figs.~\ref{Figure.1}(c) and \ref{Figure.1}(d), respectively. (d)-(e) Corresponding eigen-state distributions of two eigen-modes labeled by the red arrow in (b) and blue arrow in (c).}\label{Figure.3}
\end{figure}

We then explore properties of all modes in the projected spectrum versus $l$ in Fig.~\ref{Figure.3}(a). For $l=1$, the magnetic flux is uniformly distributed, and there is no magnetic dislocation, so no localization mode exists either. Nevertheless, for cases with $l\neq 1$, the magnetic dislocation is introduced, and various types of dislocation modes appear. In particular, one sees that the line localization modes (in blue) are widely distributed outside of the bulk spectra. The most interesting feature is that the point localization modes in red monotonically increase (decrease) with the variation of $l$ in the upper (lower) spectral gap, which gives multiple monotonic curves, and each of them shows the rainbow feature. Such rainbow distribution of point localization modes with the variation of $l$ provides a way to selectively localize a particular frequency component of the light by varying the magnetic dislocation. Moreover, the number of these monotonic curves for the point localization modes depends on the accumulated phase at the point discontinuity (i.e., the magnetic singularity) \cite{54}.

To demonstrate the magnetic singularity is the composition of two effective mechanisms, we perform separate spectrum analysis for both. In Fig.~\ref{Figure.3}(b), we plot the spectrum versus $k_q$ from the effective model described in Fig.~\ref{Figure.1}(c), again under PBC. The bands in-between bulks refer to the topological defect with the localization feature of its eigen-state distribution [see Fig.~\ref{Figure.3}(d)]. Therefore, the varying $l$ in Fig.~\ref{Figure.3}(a) effectively equals to tuning $k_q$ in the manner of $k_q\cdot d_q=-\Phi^{(3)}+\phi$ (with $d_q$ being the lattice constant in the $q$ dimension) \cite{54} which gives the rainbow feature of the point localization mode. As for the magnetic domain-wall model in Fig.~\ref{Figure.1}(d), we plot the corresponding spectrum in Fig.~\ref{Figure.3}(c) under PBC. One can see the similar spectrum with that in Fig.~\ref{Figure.3}(a) except for no point localization modes. In this effective model with the domain wall throughout the vertical range, the corresponding edge-state distribution in Fig.~\ref{Figure.3}(e) exhibits such edge mode occupying middle sites with all $n$.

\begin{figure}[htbp]
\centering
\includegraphics[width=0.9\textwidth ]{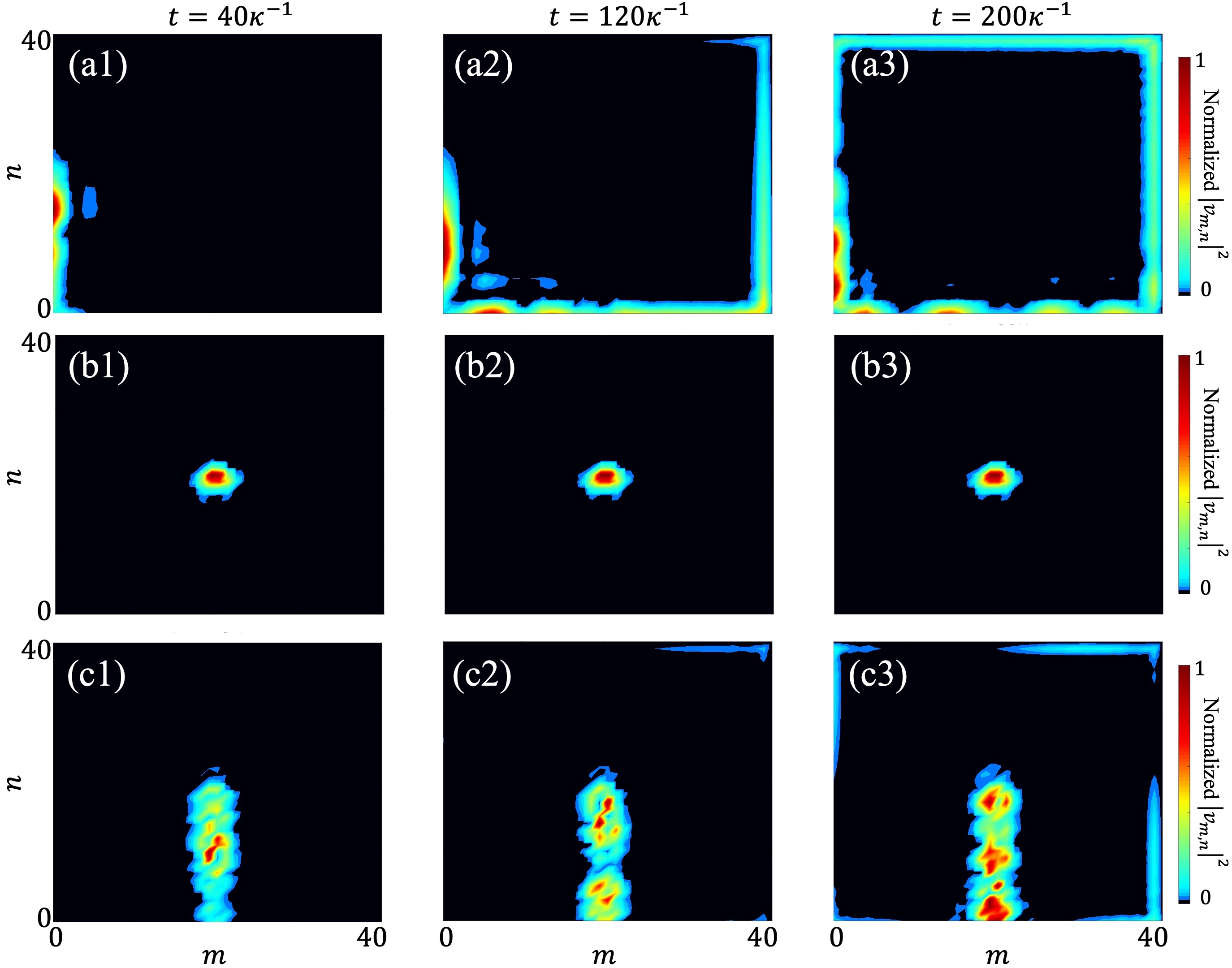}
\caption{The dynamic evolutions of the light excited by a pulse source in the finite lattice. (a) Simulation results of $|v_{m,n}|^2$ at $t=40$, $120$, $200\kappa^{-1}$, respectively, with the $E_\mathrm{in}(t)$ at $\omega_\mathrm{c}=$ 2.39 $\kappa$ excited at $(m,n)=(1,20)$. (b) Simulation results with the same $E_\mathrm{in} (t)$ at $\omega_\mathrm{c}=$ 2.39$\kappa$ but excited at $(m,n)=$ (20, 21). (c) Simulation results with the same $E_\mathrm{in} (t)$ but $\omega_\mathrm{c}=$ 1.4$\kappa$ excited at $(m,n)=$ (20,11).}\label{Figure.4}
\end{figure}

We perform simulations in a ($40\times 40$) lattice under OBC. The magnetic dislocation parameter is chosen as $l=$ 0.15, supporting point localization modes and line localization modes in-between the spectral gap. One-way edge states in-between gaps exist due to the effective magnetic flux \cite{27}. We excite the lattice at a single lattice site with the input pulse field as $E_\mathrm{in}(t)=e^{-i\omega_\mathrm{c}t}e^{-(t-t_0)^2/\Delta t^2}$, where $t_0=20\kappa^{-1}$ and $\Delta t=8\kappa ^{-1}$, and study distributions of the field intensity on each site $|v_{m,n}|^2$ at different times \cite{54}. $\omega_\mathrm{c}$ is the center frequency of the excitation.

We first excite the lattice by using $E_\mathrm{in}(t)$ with $\omega_\mathrm{c}=$ 2.39$\kappa$ at the boundary of the lattice $(m,n)=(1,20)$ to excite the topological one-way edge state. The simulation results are shown in Fig.~\ref{Figure.4}(a), where three snapshots at $t=40$, $120$, and $200\kappa^{-1}$, respectively, present the evolution of the light in the lattice at different times. The edge mode is unidirectionally propagating along the boundary of the lattice without efficiently leaking into the bulk or coupling to any localization modes.

We now take the same input pulse but excite the lattice at the point discontinuity $(m,n)=$ (20,21), and show simulation results in Fig.~\ref{Figure.4}(b). One can see that the point localization mode is successfully excited, where the light is trapped at the singularity of the magnetic dislocation, consistent with the theoretical prediction in Fig.~\ref{Figure.2}(c). More importantly, although the input at the same excitation frequency can also excite a topological edge mode if the excitation position is at the boundary of the lattice [see Fig.~\ref{Figure.4}(a)], the trapped light does not leak into the edge mode. In other words, due to the strong localization effect, the point localization mode does not interact with the edge mode due to no spatial overlap between two modes. There is no leak to the line localization mode either because of different eigen-energies for these two modes [see Fig.~\ref{Figure.2}(a)]. Moreover, the excited distribution of the point localization mode is robust against disorders (see supplementary material \cite{54}), so it showcases an efficient localization effect from the magnetic dislocation in photonics.

\begin{figure}[htbp]
\centering
\includegraphics[width=0.9\textwidth ]{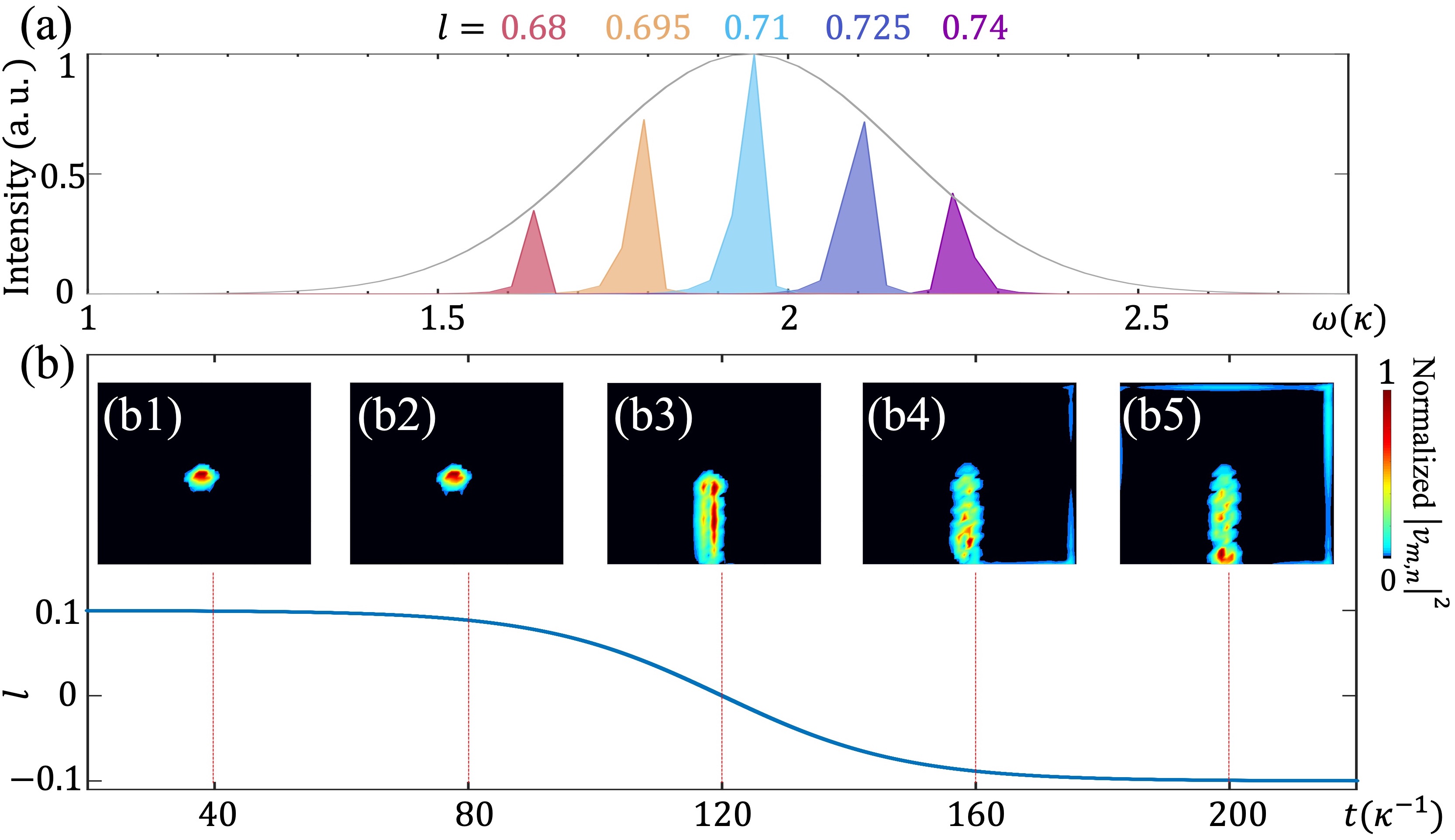}
\caption{(a) The spectra of localized modes from the field collected at $(m,n)=$($20,21$), where $l=0.68$ (red), $0.695$ (orange), $0.71$ (cyan), $0.725$ (blue), and $0.74$ (purple) are taken in simulations. The grey curve indicates the spectrum of the input pulse. (b) Adiabatic pumping of light by slowly tuning $l$ from 0.1 to $-0.1$ during the time $t\in[0,240]\kappa^{-1}$. (b1)-(b5) Simulation results of $|v_{m,n}|^2$ at $t=40$, $80$, $120$, $160$, and $200\kappa^{-1}$.}\label{Figure.5}
\end{figure}

In Fig.~\ref{Figure.4}(c), we give simulation results of exciting the lower line localization mode by using $E_\mathrm{in} (t)$ with $\omega_\mathrm{c}=$ 1.4$\kappa$ at the middle of the line discontinuity $(m,n)=$ (20, 11). Here, the light is initially trapped within the line discontinuity regime, but during the time evolution, a portion of light leaks to edge modes and propagates along the boundary counter-clockwisely. Such a phenomenon indicates that even though we successfully excite the line localization mode initially, it can interact with the one-way edge mode in a finite lattice as the small overlap between two modes, and the energy of localization is then leaking out in this case. Note that the artificial extension to top sites in Fig.~\ref{Figure.2}(b) indeed disappears.

We suggest two possible applications with the unveiled light trapping with point localization modes under manipulations of the parameter $l$. Firstly, we use a broadband pulse with $\Delta t=4.62\kappa^{-1}$, $t_0=20\kappa^{-1}$, and $\omega_\mathrm{c}=$ 1.94$\kappa$ to excite the system at the point discontinuity $(m,n)=$ (20, 21) in various choices $l$. Such pulse covers a broad spectral bandwidth $\Delta\omega\sim 0.71\kappa$, spanning over the upper spectral gap in Fig.~\ref{Figure.3}(a). We perform simulations with $l=0.68$, $0.695$, $0.71$, $0.725$, and $0.74$, respectively, collect field amplitudes at $(m,n)=$ (20, 21), and plot the corresponding spectra in Fig.~\ref{Figure.5}(a). One can see the selective excitation of the localization mode with the blue shift of the frequency when $l$ is increasing.

Next, we show the adiabatic pumping of light by exciting the system at (20, 21) with pulse parameters $\omega_c=1.96\kappa$, $t_0=20\kappa^{-1}$, and $\Delta t=8\kappa^{-1}$, and then slowly varying $l$ from 0.1 to $-0.1$ shown in Fig.~\ref{Figure.5}(b). The system is initially excited at the point localization mode and then adiabatically gets tuned into the regime of line localization modes [see Fig.~\ref{Figure.3}(a)]. The simulation results show that the field is initially trapped at the middle of the lattice for $t=40\kappa^{-1}$, $80\kappa^{-1}$, but gets converted into the line-localization distribution at $t=120\kappa^{-1}$. Afterwards, the field slowly leaks into the edge mode. This example exhibits interesting conversion among three modes, namely point localization, line localization, and edge state modes with an adiabatic operation, showing a way for releasing the trapped light at the magnetic singularity in a controllable way.

\section{Summary}
We study the magnetic dislocation in a spatially uniform lattice and explore physics of localization modes in theory. Due to the singularity in the effective magnetic field, light can be trapped robustly in the vicinity of the singularity without leaking into either bulk modes or edge modes. We find the rainbow featured light-trapping phenomena, which supports possible applications for selectively excitation of the localization mode and adiabatically pumping light out of such trapped state. Our research explores a universal Hamiltonian in photonics, where the desired hopping phases for magnetic dislocation may be possibly realized in many experiment-feasible platforms other than the proposed synthetic frequency lattice model \cite{54}. For example, synthetic modal space where hopping phases can be designed between propagating modes in curved waveguide arrays \cite{32} can be another reliable candidate. For spatial photonics platforms, one can use ring arrays with designed auxiliary delayed lines for realizing desired hopping phases \cite{26}. Moreover, the magnetic dislocation may also be generalized to other 2D symmetric lattices, for example, the C$_6$ symmetric lattice \cite{54}.

\begin{acknowledgments}
The research was supported by the National Natural Science Foundation of China (12122407, 11974245, 12192252, 12174072, and 2021hwyq05), National Key Research and Development Program of China (2023YFA1407200 and 2021YFA1400900). L.Y. thanks the sponsorship from Yangyang Development Fund.
\end{acknowledgments}


\end{document}